\documentclass[final,3p,times]{elsarticle}

\usepackage{amssymb}    
\usepackage{amsmath}    
\usepackage{lipsum}  
\usepackage{graphicx}   
\usepackage{subcaption}
\usepackage{accents}    
\usepackage{array}      
\usepackage{color}      
\usepackage{hyperref}
\usepackage{float}
\hypersetup{
  pdftitle={MRI-MECH: Mechanics-informed MRI to estimate esophageal health},
  pdfauthor={Sourav Halder}
}

\journal{ArXiv}

\begin{document}

\begin{frontmatter}

\title{MRI-MECH: Mechanics-informed MRI to estimate esophageal health}

\author{Sourav Halder\fnref{LB1}}
\author{Ethan M. Johnson\fnref{LB2}}
\author{Jun Yamasaki\fnref{LB3}}
\author{Peter J. Kahrilas\fnref{LB4}}
\author{Michael Markl\fnref{LB2}}
\author{John E. Pandolfino\fnref{LB4}}
\author{Neelesh A. Patankar\corref{cpau}\fnref{LB1,LB3}}
\cortext[cpau]{Corresponding author}
\ead{n-patankar@northwestern.edu}

\address[LB1]{Theoretical and Applied Mechanics Program, Northwestern University, Evanston, IL 60208, USA}
\address[LB2]{Department of Radiology, Northwestern University, Chicago, IL 60611, USA}
\address[LB3]{Department of Mechanical Engineering, Northwestern University, Evanston, IL 60208, USA}
\address[LB4]{Division of Gastroenterology and Hepatology, Feinberg School of Medicine, Northwestern University, Chicago, IL 60611, USA}

\begin{abstract}
Dynamic magnetic resonance imaging (MRI) is a popular medical imaging technique to generate image sequences of the flow of a contrast material inside tissues and organs. However, its application to imaging bolus movement through the esophagus has only been demonstrated in few feasibility studies and is relatively unexplored. In this work, we present a computational framework called mechanics-informed MRI (MRI-MECH) that enhances that capability thereby increasing the applicability of dynamic MRI for diagnosing esophageal disorders. Pineapple juice was used as the swallowed contrast material for the dynamic MRI and the MRI image sequence was used as input to the MRI-MECH. The MRI-MECH modeled the esophagus as a flexible one-dimensional tube and the elastic tube walls followed a linear tube law. Flow through the esophagus was then governed by one-dimensional mass and momentum conservation equations. These equations were solved using a physics-informed neural network (PINN). The PINN minimized the difference between the measurements from the MRI and model predictions ensuring that the physics of the fluid flow problem was always followed. MRI-MECH calculated the fluid velocity and pressure during esophageal transit and estimated the mechanical health of the esophagus by calculating wall stiffness and active relaxation. Additionally, MRI-MECH predicted missing information about the lower esophageal sphincter during the emptying process, demonstrating its applicability to scenarios with missing data or poor image resolution. In addition to potentially improving clinical decisions based on quantitative estimates of the mechanical health of the esophagus, MRI-MECH can also be enhanced for application to other medical imaging modalities to enhance their functionality as well.
\end{abstract}

\begin{keyword}
MRI \sep esophagus \sep physics-informed neural network \sep computational fluid dynamics \sep deep learning \sep lower esophageal sphincter \sep active relaxation \sep esophageal wall properties
\end{keyword}

\end{frontmatter}

\section{Introduction} \label{sec:intro}
The esophagus plays a crucial role in the functioning of the gastrointestinal tract and esophageal disorders are associated with reduced quality of life. There is a high worldwide prevalence of esophageal disorders exemplified by a studies \cite{Serag871,yamasaki2018} reporting that gastro-esophageal reflux disease (GERD) has a prevalence of $18.1\%-27.8\%$ in North America alone with an increase across all age groups. Another study \cite{bhattacharyya2014} reported that dysphagia (swallowing difficulty) affects 1 in 25 adults annually in the United States. Hence, it is important to improve current diagnostic technologies for esophageal disorders. Some of the common tests for diagnosing esophageal disorders are barium esophagram using fluoroscopy, high resolution manometry (HRM) \cite{fox2004,pandolfino2007,pandolfino2008,fox2008,pandolfino2009}, and functional lumen imaging probe (FLIP) \cite{gyawali2013,carlson2016}. An esophagram is non-invasive test wherein a patient swallows a radiopaque material, usually dilute barium, and fluoroscopic imaging is used to visualize the esophageal lumen. HRM and FLIP are more invasive procedures where a catheter with sensors is inserted into the esophagus to quantitatively assess the esophageal contractility. Measurements made by HRM and FLIP are physical quantities such as the pressure developed within the esophagus when a fluid is swallowed and/or the cross-sectional area variation along the esophageal length. Variations of these physical quantities are the consequence of more fundamental esophageal physiomarkers such as the stiffness of the esophageal walls, active contraction of esophageal musculature, and active relaxation. However, clinical decisions are made based on the qualitative or quantitative patterns of these physical quantities rather than the physiomarkers that cause them. For example, the widely used Chicago Classification v4.0 (CCv4.0) \cite{chicago_classification} classifies esophageal disorders based on a set of parameters derived from pressure measurements made with HRM. The explanation for this is that it is difficult to measure the fundamental physiomarkers which occur at molecular, cellular, and tissue levels. Since luminal pressure and cross-sectional area, which occur at a tissue level, are the physical quantities commonly measured by HRM and FLIP, the first stage of quantifying the fundamental physiomarkers of esophageal function are at the tissue level. In this context, the mechanical properties of the esophageal wall as well as its dynamic behavior related to active contraction and relaxation could be important physiomarkers. Thus, mechanics-based analysis may provide valuable mechanistic insight regarding esophageal function.

Previous mechanics-based studies on the esophagus have been done both experimentally and computationally. Experimental studies \cite{fan2004,yang2006,natali2009,sokolis2009,sokolis2013}  focused on the mechanical properties of the esophageal walls in-vitro. In-silico modeling of the esophagus have been performed both in the context of pure fluid mechanics \cite{brasseur1987,li1993,li1994,ghosh2005,acharya_peristalsis} to understand the nature of bolus transport as well as fully resolved fluid-structure interaction models to understand how the esophageal muscle architecture influences esophageal transport as well as the stresses developed in the esophageal walls during bolus transport \cite{kou2015,kou2017}. In-silico mechanics-based analysis have also been performed on data obtained from the various diagnostic devices to identify mechanics-based physiomarkers. Acharya et al. \cite{acharya_egjw} used a mechanics-based approach to calculate the work done by the esophagus in opening the esophagogastric junction (EGJ) and the necessary work required to open the EGJ using data obtained from the FLIP. Halder et al. \cite{halder2021} introduced a framework called FluoroMech applied to fluoroscopy images to estimate the mechanical health of the esophagus. FluoroMech enhances the capability of fluoroscopy by adding quantitative predictions to fluoroscopy data which is inherently qualitative in nature. In this work, we present a framework called MRI-MECH which uses dynamic MRI as input to estimate esophageal health. 

Both FluoroMech and MRI-MECH utilize the input of esophageal cross-sectional area varying as a function of time and length along the esophagus. However, there are some key differences in their approach that can be classified into two categories. The first category pertains to differences between fluoroscopy and dynamic MRI. Fluoroscopy is an older and simpler approach wherein X-ray imaging is used to visualize a swallowed bolus passing through the esophagus resulting in a video with high temporal resolution, but only a two-dimensional projection of the bolus. Hence, the three-dimensional geometry of the bolus is unknown. Fluoroscopy is a well established clinical test. Dynamic MR imaging, on the other hand, is a relatively complicated and evolving technology. In its current state, dynamic MRI images have significantly lower temporal resolution but very detailed three-dimensional representation of the bolus. However, dynamic MR imaging is currently not a standard practice for evaluating esophageal disorders offering a vast potential for improvement. The second category of differences between FluoroMech and MRI-MECH lie in the implementations of the frameworks. FluoroMech uses the finite volume method to predict esophageal wall stiffness and active relaxation with the variation of cross-sectional area as input. It is computationally fast (less than a minute) and requires very limited computational resources but requires a complete dataset of the variation of cross-sectional area. Assumptions are required regarding the 3-D shape of the bolus based on the volume of fluid swallowed and since model predictions are sensitive to cross-sectional area variation, inaccuracies in measurements reflect on the predictions as well. MRI-MECH, on the other hand, uses a physics-informed neural network (PINN) \cite{RAISSI2019} to make predictions and is much computationally demanding (takes approximately one hour to run) requiring better hardware, especially the GPU, to train the PINN. However, MRI-MECH is not sensitive to missing or imperfect measurements. Additionally, it does not require assumptions regarding the esophageal lumen cross-sectional shape because MRI provides three-dimensional geometry of the esophageal lumen. In the following sections, we describe the MRI-MECH framework in detail along with its application to a dynamic MRI sequence.
\begin{figure}
  \centerline{\includegraphics[scale=0.15]{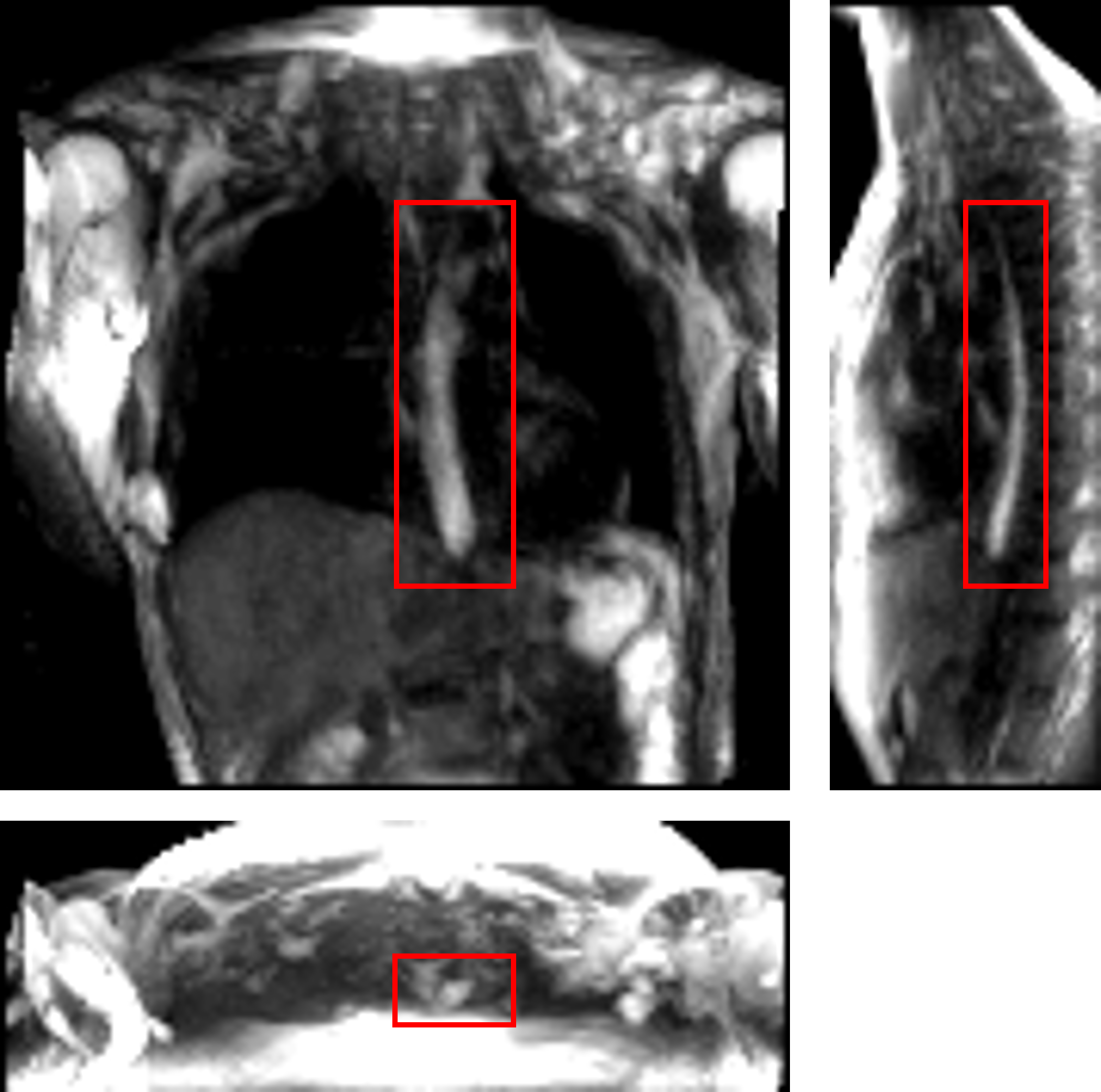}}
  \caption{One instant of a dynamic MRI of a normal subject as seen in three perpendicular planes. The planes (from left to right to bottom) are coronal, sagittal and axial, respectively. The bolus can be seen as the bright region inside the red boxes. Concentrated pineapple juice was swallowed as a contrast agent.}
  \label{fig-mri}
\end{figure}  

\section{Accelerated dynamic MRI}
Imaging was performed at 1.5 T (Aera, Siemens, Germany) using a 3D MR angiography sequence (TWIST, Siemens, Germany) designed for contrast-enhanced cardiac imaging applications which was adapted to be used for esophageal imaging using pineapple juice as an oral contrast agent. Sequence parameters included (3.25 $\mathrm{mm})^3$ spatial / 1.17 s temporal resolution, (416 $\mathrm{mm})^2\times143$ $\mathrm{mm}$ coronal field of view, 0.78 ms echo time, 2.36 ms repetition time, $29^\mathrm
{o}$ tip angle, 620 Hz/pixel bandwidth, 6/8 partial Fourier acquisition, R=2 GRAPPA acceleration, $8\%$ central size / $10\%$ outer density view sharing. A 4-channel cardiac coil was used for image acquisition, placed on the upper torso surface. To improve image conspicuity of the juice bolus, pineapple juice (100$\%$, Costa Rica) was reduced to a volume factor of 0.48 (i.e. 52$\%$ volume removed) through gradual heating without boiling. By doing so, the T1 of the juice at 1.5 T was reduced from 265 ms (raw / non-volume reduced juice) to 76 ms (volume-reduced juice), as measured by variable flip angle signal fit. A healthy volunteer (37 year old male) was given 20 ml of the volume-reduced pineapple juice to swallow during image acquisition. The juice was administered via a plastic tube and syringe controlled by the scan subject. The subject was instructed to swallow by voice command from the scan operator, given 10 seconds after the start of image acquisition, with 75 seconds of imaging performed to capture complete esophageal transit. To visualize the bolus transport, maximum intensity projections were created. Figure \ref{fig-mri} shows an instant during bolus transport on three perpendicular slices. 
\section{Extraction of bolus geometry}
\begin{figure}[h]
  \centerline{\includegraphics[width=\textwidth]{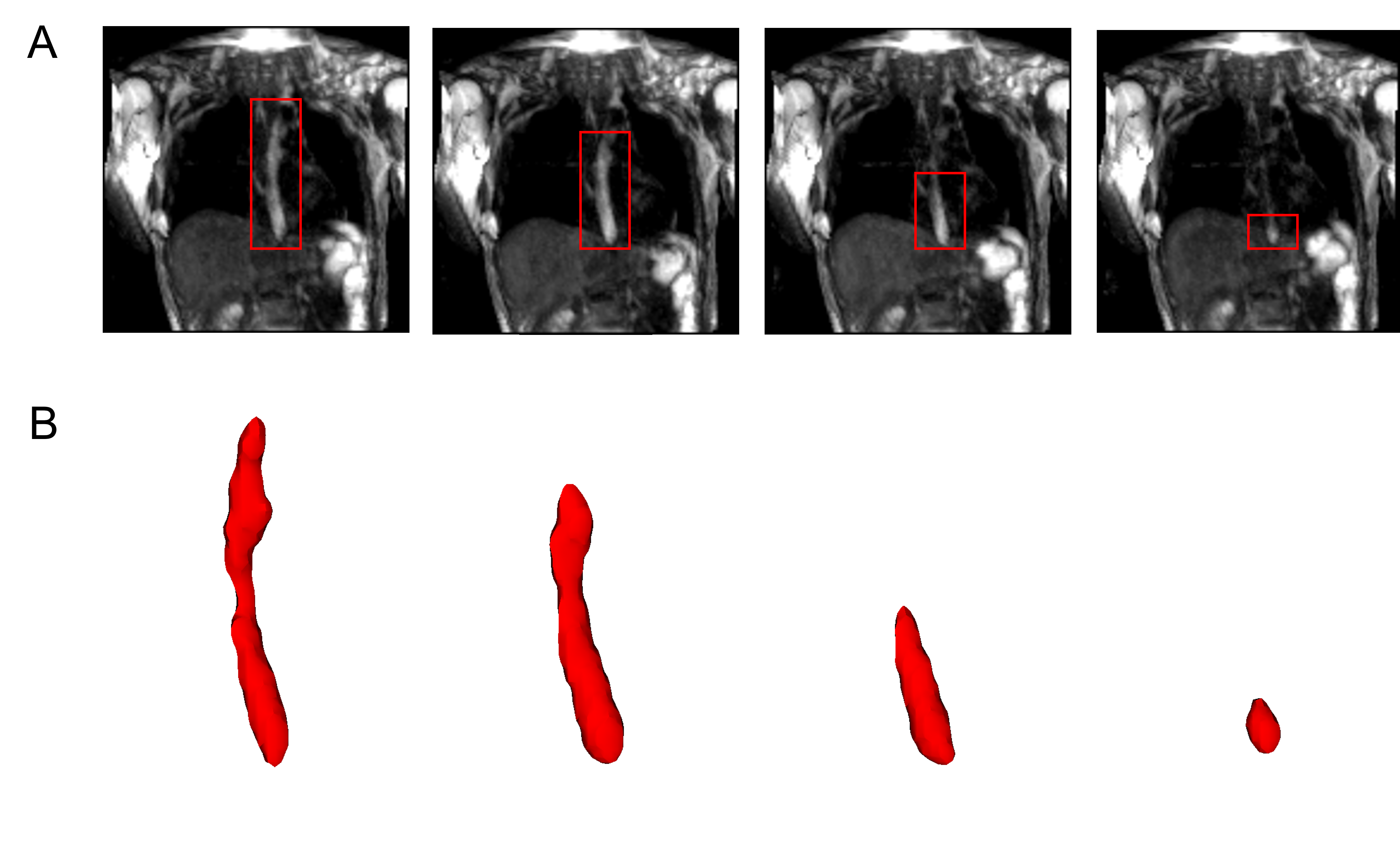}}
  \caption{Segmentation of MR images. (A) The bolus is shown in the coronal plane at four time instants (progressing from left to right). The bolus is seen as the bright region inside the red boxes. The bolus volume decreased with time as it was emptied into the stomach. (B) The corresponding 3D segmented bolus shapes for the four time instants. Bolus size has been magnified for visualization.}
  \label{fig-segmentation}
\end{figure}
The MRI output consisted of a cuboid wherein voxels in a Cartesian coordinate system had different magnitudes of intensity. The temporal resolution of the dynamic MRI (1.17 second) determined the number of images with the bolus seen within the esophagus; 7 time instants in this study. The typical length of an adult esophagus is 18 - 25 cm \cite{Oezcelik2011}. The average velocity of a normal peristalsis is approximately 3.3 cm/s \cite{hollis1975}. Thus, an average swallow sequence usually takes 5 - 8 seconds. Therefore, temporal resolutions similar to what we used in our analysis typically result in 5 - 8 images. Although this temporal resolution is not comparable to fluoroscopy, the detailed three-dimensional geometry of the bolus in MRI leads to better prediction of velocity and intrabolus pressure resulting in better prediction of esophageal wall properties. The bolus was manually segmented for the 7 time instants, a few of which are shown in Figure \ref{fig-segmentation}. The segmentation assigned a value of 1 and 0 to each voxel that lay inside and outside the bolus, respectively. The image segmentation was performed using the open-source software ITK-SNAP \cite{itksnap}. With improved MR imaging and better temporal resolution, manual image segmentation might not be feasible and more sophisticated automated segmentation techniques might be necessary. We have described a deep learning based automated segmentation approach called 3D-U-Net \cite{3dunet} in the Appendix which was fine-tuned for this application.

\begin{figure}[h]
  \centerline{\includegraphics[scale=0.15]{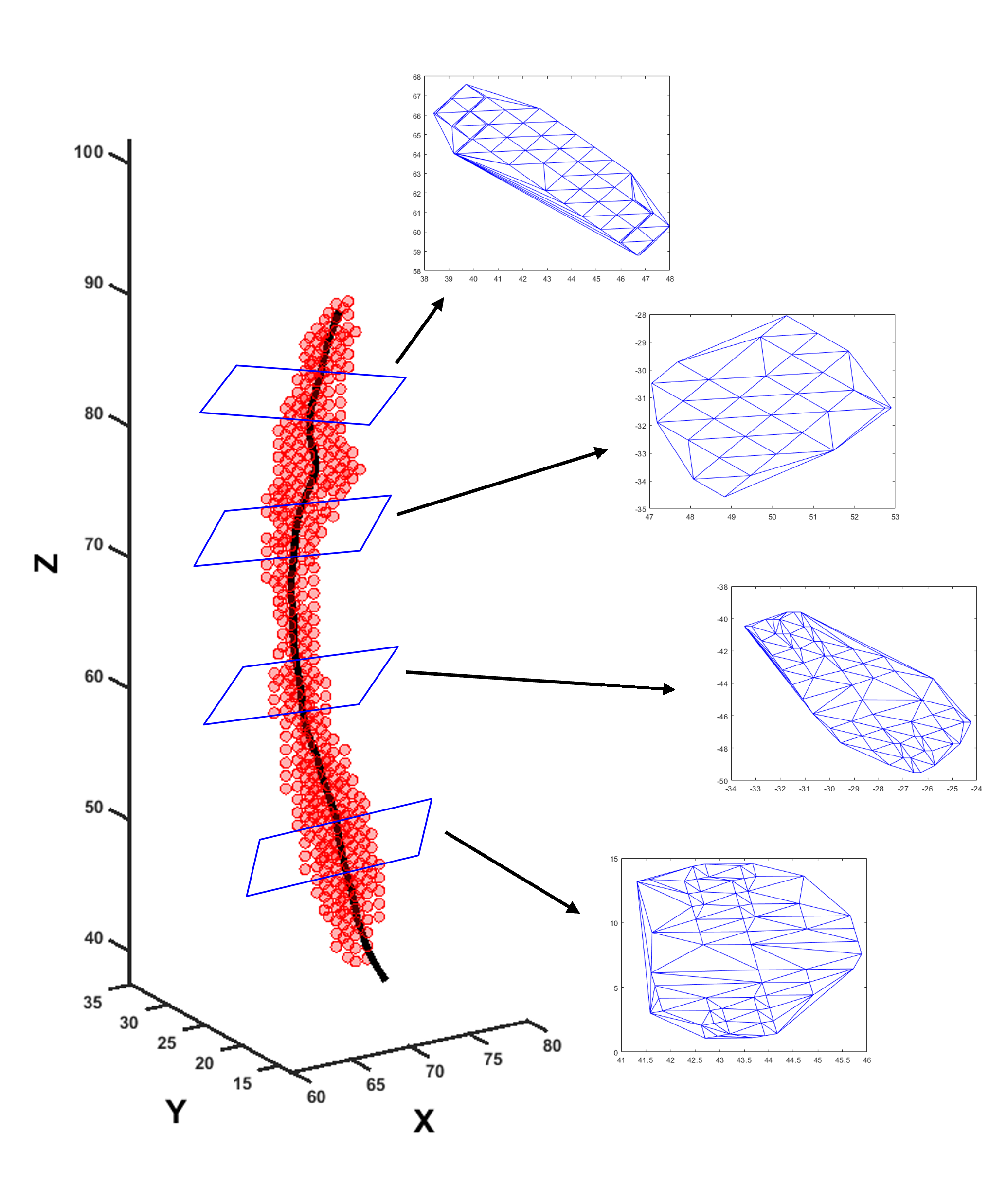}}
  \caption{Extraction of cross-sectional areas from dynamic MR images. The segmented bolus geometry at one time instant is shown by the red points in the scatter plot. The generated center line is shown by the black curve inside. A few planes are shown which are perpendicular to the center line and on which the cross-sectional areas were calculated. The points on the planes were meshed using Delaunay triangulation and the triangulated shapes approximate the cross-sectional areas at those planes.}
  \label{fig-extract_csa}
\end{figure}

MRI-MECH modeled the esophagus as a one-dimensional flexible tube. For such one-dimensional analysis, the variation of cross-sectional areas at different points along the length of the esophagus and different time instants had to be extracted from the three-dimensional bolus obtained from segmentation. This was done in two steps. The first step was to generate a center line along the length of the esophagus. The bolus shapes observed at different time instants were superimposed and then cross-sections of the superimposed shape at different horizontal planes from the proximal to the distal end of the superimposed shape were generated. The centroids of these cross-sections were connected to form the center line. The length of the center line in this case was 9.65 cm. The second step, after extracting the center line, was to generate planes perpendicular to the centerline as shown in Figure \ref{fig-extract_csa}. The segmented voxels marked 1 which lay near these perpendicular planes were projected onto these planes. These projected points were connected using Delaunay triangulation as shown in Figure  \ref{fig-extract_csa}. The cross-sectional area at each point along the center line was then calculated as the sum of the triangles in the Delaunay triangulated geometries.
\section{MRI-MECH formulation}
\subsection{Governing equations}
Transport through the esophagus was modeled as one-dimensional fluid flow through a flexible tube. The mass and momentum conservation equations in one dimension are as follows \cite{barnard1966,kamm1979,manopoulos2006,ottesen2006}:
\begin{align}
 \frac{\partial A}{\partial t} + \frac{\partial (AU)}{\partial x} =0, \label{eqn-continuity} \\
 \frac{\partial U}{\partial t} + \frac{\partial}{\partial x}\left(\frac{U^2}{2}\right)+\frac{1}{\rho}\frac{\partial P}{\partial x} + \frac{8\pi\mu U}{\rho A} = 0, \label{eqn-momentum}
\end{align}
where $A$ is the cross-sectional area of the esophageal lumen,  $U$ and $P$ are the velocity and pressure in the bolus fluid, respectively. $x$ represents the distance along the length of the esophagus from the mouth to the stomach and $t$ represents time. The total time for bolus transport in our analysis was 6.95 seconds. $\rho$ and $\mu$ are the density and dynamic viscosity of the transported fluid, respectively. Pineapple juice was the swallowed fluid whose density and viscosity were 1.06 g/cm$^3$ and 0.003 Pa.s \cite{shamsudin2009}, respectively.

It has been observed experimentally that the fluid pressure developed inside the esophagus is linearly proportional to the cross-sectional area of the esophageal lumen \cite{orvar1993,kwiatek2010} in the absence of any neuromuscular activation. Using this information, a pressure tube law can be constructed as follows:
\begin{align}
P=P_o+K\left(\frac{A}{\theta A_o}-1\right), \label{eqn-tube_law}
\end{align}
where $K$ is the stiffness of the esophageal wall, $P_o$ is the pressure outside the esophageal wall and is often equal to the thoracic pressure, $A_o$ is the cross-sectional area of the esophageal lumen in its inactive state, and $\theta$ is the activation parameter. Typically, the inactive cross-sectional area is in the range 7-59 $\mathrm{mm}^2$ \cite{xia2009}. In this case, the inactive cross-sectional area $A_o$ was $27 \mathrm{ mm}^2$. The activation parameter $\theta$ takes the value of 1 in the inactive state of the esophagus. It can be seen from Equation \ref{eqn-tube_law} that in the inactive state, when the cross-sectional area of the esophageal lumen is equal to $A_o$, the pressure inside the esophagus is equal to $P_o$. Due to the lack of information about the thoracic pressure, we assume that $P_o=0$ mmHg. An activation is induced when $\theta<1$ raising the pressure locally. On the other hand, $\theta>1$ decreases the bolus pressure and estimates the active relaxation of the esophageal wall. Thus, the parameter $\theta$ captures the effect of the esophageal motility. 

Due to the low resolution of the dynamic MRI, it was necessary to interpolate the MRI data to smaller temporal and spatial scale. The measured volume $V_m$ of the bolus from the proximal end ($x=0$) to any point $x>0$ was calculated as follows:
\begin{align}
V_m = \int_0^x A_m dx, \label{eqn-volume}
\end{align}    
where $A_m$ is the measured cross-sectional area of the esophageal lumen at a coarse $x$ and $t$. The volume $V_m$ was interpolated using piecewise cubic Hermite interpolating polynomial to a smaller temporal and spatial scale to obtain $V$. $V_o$ was known at 7 time instants and 59 points along $x$. The interpolated $V$ was calculated at 100 time instants and 100 points along $x$. Using Equations \ref{eqn-continuity} and \ref{eqn-volume}, the cross-sectional areas and velocities at finer $t$ and $x$ were calculated as follows:
\begin{align}
A &= \frac{\partial V}{\partial x}, \label{eqn-csa} \\
U &= -\frac{1}{A}\frac{\partial V}{\partial t}. \label{eqn-velocity}
\end{align}
The values of $A$ and $U$ were then used to solve for $P$ in Equation \ref{eqn-momentum}. Equations \ref{eqn-continuity} and \ref{eqn-momentum} were non-dimensionalized as follows:
\begin{align}
 \frac{\partial \alpha}{\partial \tau} + \frac{\partial (\alpha u)}{\partial \chi} &=0, \label{eqn-continuity_nd} \\
 \frac{\partial u}{\partial \tau} + \frac{\partial}{\partial \chi}\left(\frac{u^2}{2}\right)+ \frac{\partial p}{\partial \chi} + \varphi\frac{u}{\alpha} &= 0, \label{eqn-momentum_nd}
\end{align}
where $\alpha=A/A_s$, $u=U/c$, $\chi=x/\sqrt{A_s}$, $\tau=ct/\sqrt{A_s}$, $p=P/(\rho c^2)$, $\varphi=(8\pi\mu)/(\rho c \sqrt{A_s})$, $A_s=\mathrm{max}(A)$, and $c=5$ cm/s is a reference speed of peristalsis. In this work, $A_s=197.73$ $\mathrm{mm}^2$. Using the properties of the swallowed fluid and the scales for $A$ and $U$, we found $\varphi= 0.101$. The pressure tube law as described in Equation \ref{eqn-tube_law} was non-dimensionalized as follows:
\begin{align}
 p &= k\left(\frac{\alpha}{\theta\alpha_o} - 1\right), \label{eqn-tube_law_nd}
\end{align} 
where $k=K/(\rho c^2)$ and $\alpha_o=A_o/A_s$. This non-dimensionalization ensures that the magnitudes of $\alpha$, $u$, and $p$ lie between -1 and 1, which is essential for good prediction by the PINN as described later.

\subsection{Initial and boundary conditions}
 \begin{figure}[h]
  \centerline{\includegraphics[scale=0.5]{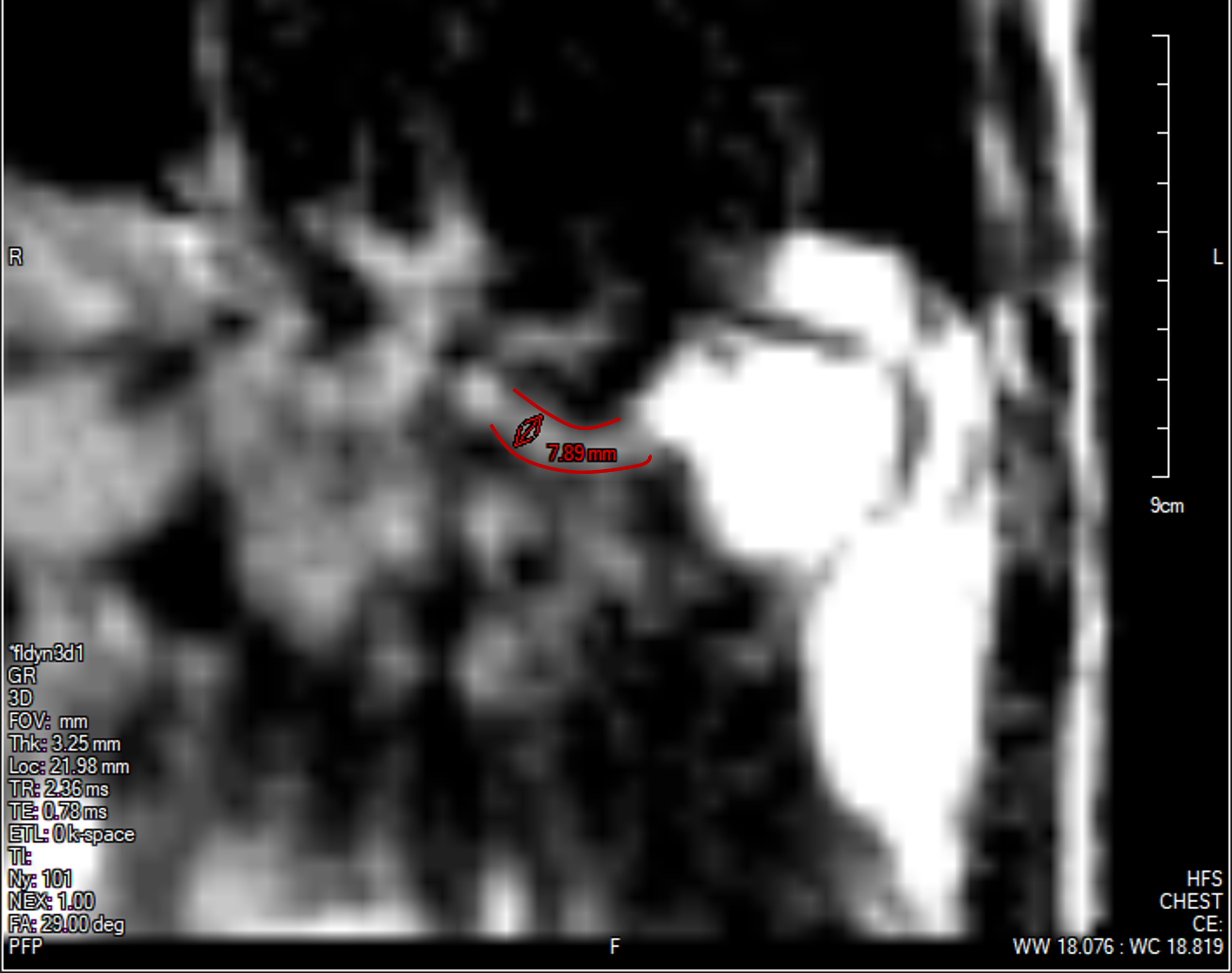}}
  \caption{The lower esophageal sphincter (LES) identified at a single time instant outlined in red with a diameter of approximately 7.89 mm and a length of approximately 2.78 cm. The stomach can be seen to the right of the LES with the accumulated pineapple juice shown in bright white. The esophageal body cannot be seen in this slice because this plane does not intersect the esophagus.}
  \label{fig-les}
\end{figure}
The boundary conditions of this problem were specified to capture the physiological conditions of normal esophageal transport. The upper esophageal sphincter (UES) at the proximal end of the esophagus opens to allow the bolus into the esophagus, closes once the fluid has passed through it, remains closed thereafter. Hence, we specified zero velocity at $x=0$ for all time instants. This condition also ensures that $V_m=0$ at $x=0$ at all time instants and is consistent with Equations \ref{eqn-volume} and \ref{eqn-velocity}. The distal end of the esophagus, on the other hand, remains open to allow emptying into the stomach. Since the pressure term in Equation \ref{eqn-momentum} consists of a single derivative with respect to $x$, it is necessary to specify only one boundary condition for $P$. The boundary pressure was specified at the distal end which was equal to a typical value of gastric pressure (7 mmHg). Finally, for initial condition, we had zero velocity at all points along $x$ at $t=0$.

\subsection{Cross-sectional area of the lower esophageal sphincter}
The low spatial resolution of the dynamic MRI poses a problem of accurately identifying the lower esophageal sphincter (LES) cross-section. This is because the LES opening is narrower compared to the esophageal body and does not distend very much because of the greater wall stiffness at the esophagogastric junction (EGJ). Although this could be improved by focusing the MRI only at the LES, the state of the esophagus proximal to the LES cannot be estimated in such a scenario. The LES can be identified in only one or two time instants when the LES has greatly distended due to a bolus flow through it. Figure \ref{fig-les} shows the LES at one such time instant. The LES cross-sectional area measured at this time instant can act as a valuable reference to identify the bolus behavior proximal to the LES. 
\begin{figure}[h]
  \centerline{\includegraphics[scale=0.6]{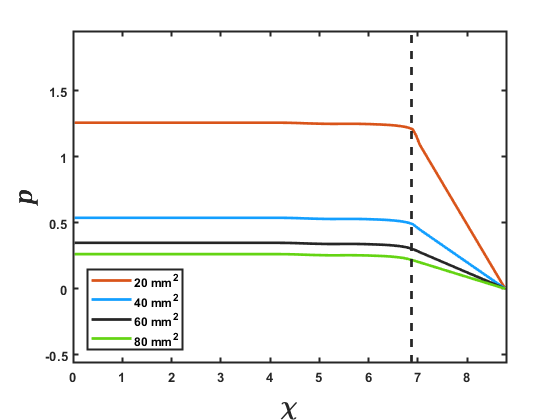}}
  \caption{The effect of LES cross-sectional area on the prediction of intrabolus pressure. The proximal end of the LES is marked by the vertical dashed line. The inserted legend shows the LES cross-sectional areas used for this simulation. Equations \ref{eqn-continuity_nd} and \ref{eqn-momentum_nd} were solved using the method described in FluoroMech \cite{halder2021} to calculate the intrabolus pressure. The input for the model was the variation of $\alpha$ observed from the MRI with four reference LES cross-sectional areas. The variation of pressure is shown at a single time instant to illustrate the impact of LES cross-sectional area on pressure prediction.}
  \label{fig-les_effect}
\end{figure}

As specified in the previous section, since pressure is specified as a Dirichlet boundary condition at the distal end of the esophagus, the intrabolus pressure prediction depends on the accurate measurement of the LES cross-sectional area. Figure \ref{fig-les_effect} shows the intrabolus pressure calculated using the numerical approach described in Halder et al. \cite{halder2021} with different LES cross-sectional areas. The pressure shown is non-dimensional and the pressure at the distal end was specified zero as a reference in this case. The total length of the esophagus considered here is the sum of the centerline length (9.65 cm) and the LES length (2.78 cm). Thus, the proximal and distal location of the bolus were 9.65 cm and 12.43 cm, respectively. In non-dimensional form, the proximal and distal locations were $\chi_p=6.87$ and $\chi_L=8.81$, respectively. The quantities $\chi_p$ and $\chi_L$ were important locations as described in the next section. As shown in Figure \ref{fig-les_effect}, the intrabolus pressure proximal to the LES depends on the LES cross-sectional area, so, assuming a constant LES cross-sectional area (measured at one time instant) would lead to an incorrect prediction, making it important to know the instantaneous LES cross-sectional area to accurately predict intrabolus as well as to understand LES functioning during emptying. 

\subsection{Physics-informed neural network}
\begin{figure}[h]
  \centerline{\includegraphics[width=\textwidth]{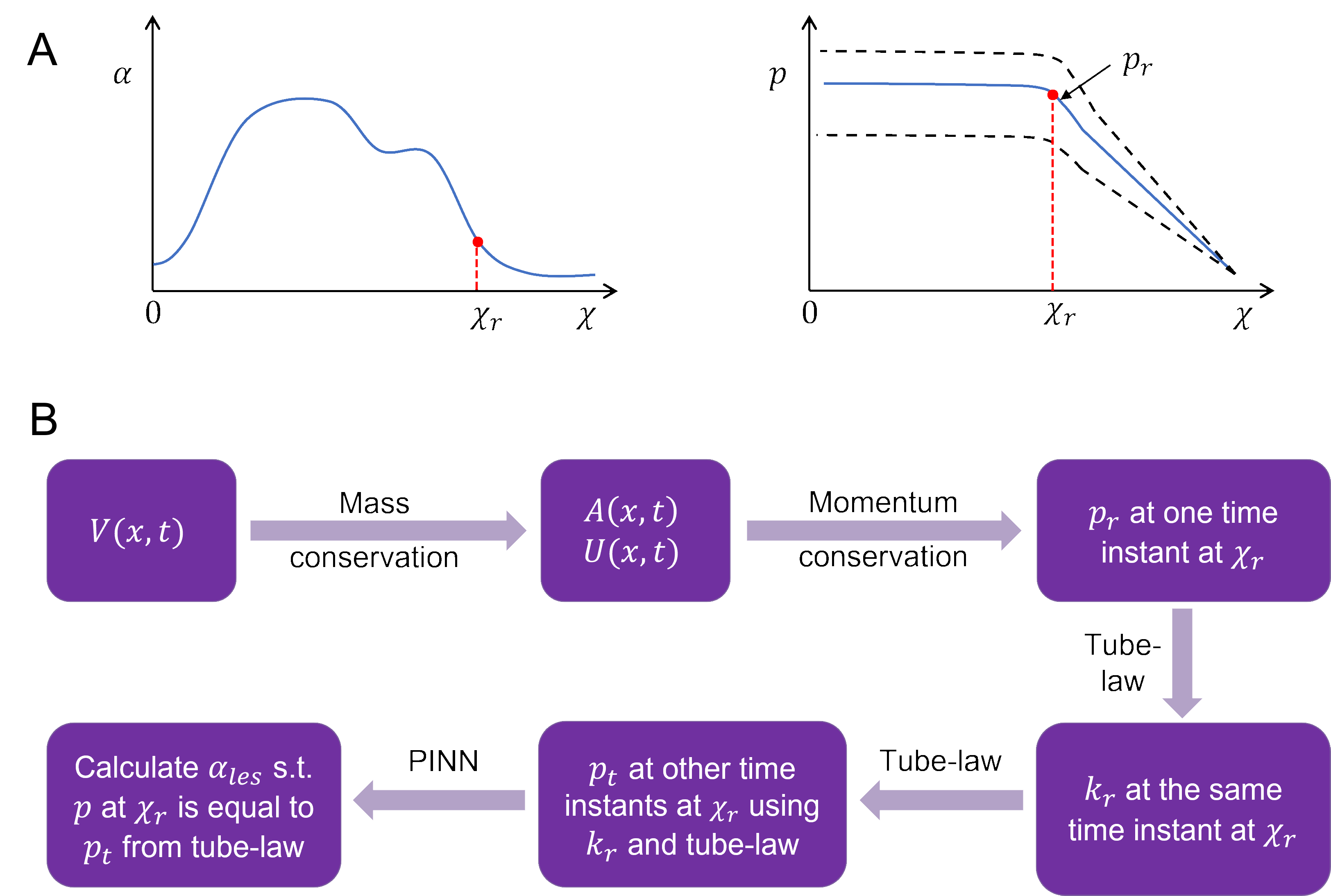}}
  \caption{Problem definition for the physics informed neural network (PINN) framework. (A) Schematic for the variation of cross-sectional area and pressure at the time instant when the LES cross-section was known. The dashed lines in the pressure variation show what intrabolus pressure would be at other time instants assuming constant LES cross-sectional area. (B) Workflow for the prediction of LES cross-sectional area at other time instants.}
  \label{fig-problem_definition}
\end{figure}
The problem of missing data for the LES cross-sectional area (and consequently obtaining accurate intrabolus pressure values) was solved using a physics-informed neural network (PINN) \cite{RAISSI2019}. The problem description is schematically shown in Figure \ref{fig-problem_definition}. The final interpolated volume $V(x,t)$ was used to calculate $A(x,t)$ and $U(x,t)$ using Equations \ref{eqn-csa} and \ref{eqn-velocity}, and after non-dimensionalization, $\alpha(\chi,\tau)$ and $u(\chi,\tau)$, respectively. These values of $\alpha(\chi,\tau)$ and $u(\chi,\tau)$ were then used to calculate $p(\chi_r,\tau_r)$ at the specific time instant when the LES cross-section was visible by solving Equation \ref{eqn-momentum_nd} using the finite volume method described in Halder et al. \cite{halder2021}. The non-dimensional time $\tau_r$ corresponds to the time instant when the LES was visible in MRI. The point $\chi_r$ was selected near the proximal end of the LES. This point was selected because the pressure at points proximal to $\chi_r$ are of similar magnitude as $p(\chi_r,\tau)$ as shown in Figure \ref{fig-les_effect}. Additionally, $\chi_r$ was very close to the LES and hence the effect of active relaxation as observed in the esophageal body was minimal. Note that this was an assumption that we made regarding the active relaxation, and its usefulness will be explained shortly. The values of $\chi_r$ and $\tau_r$ were 6.76 and 8.57, respectively. The pressure $p(\chi_r,\tau_r)$  was the correct estimate of the intrabolus pressure since the LES cross-sectional area was accurately known. We call this pressure the reference pressure,  $p_r=p(\chi_r,\tau_r)$. Using the tube law in Equation \ref{eqn-tube_law_nd}, the stiffness ($k_r$) at $\chi_r$ was calculated as follows:
\begin{align}
k_r = \frac{p_r}{\left(\frac{\alpha(\chi_r,\tau_r)}{\alpha_o}-1\right)}. \label{eqn-ref_stiffness}
\end{align}
Note that there is no $\theta$ in Equation \ref{eqn-ref_stiffness} since we assumed that $\theta=1$ at $\chi_r$. With the stiffness at $\chi_r$ known, we calculated the pressure $p_t=p(\chi_r,\tau)$ at other times with the tube law according to Equation \ref{eqn-tube_law_nd} as follows:
\begin{align}
p_{t} = k_r\left(\frac{\alpha(\chi_r,\tau_r)}{\alpha_o}-1\right). \label{eqn-ref_pressure}
\end{align}
The LES cross-sectional area ($A_{les}$) was calculated using PINN so that the pressure predicted at $\chi_r$ matches $p_t$ for all times. An additional constraint is necessary to ensure an unique solution for $A_{les}$ as follows:
\begin{align}
\frac{\partial \alpha_{les}}{\partial \chi} = 0, \label{eqn-les_constraint}
\end{align} 
where $\alpha_{les}=A_{les}/A_s$ is the non-dimensional cross-sectional area of the LES. Equation \ref{eqn-les_constraint} implies that there was no significant variation of LES cross-sectional area along $\chi$. This is physically meaningful since the variation of $\alpha_{les}$ along $\chi$ is quite negligible compared to the esophageal body and can be observed in Figure \ref{fig-les} as well.

\subsubsection{Network architecture}
\begin{figure}[h]
  \centerline{\includegraphics[scale=0.7]{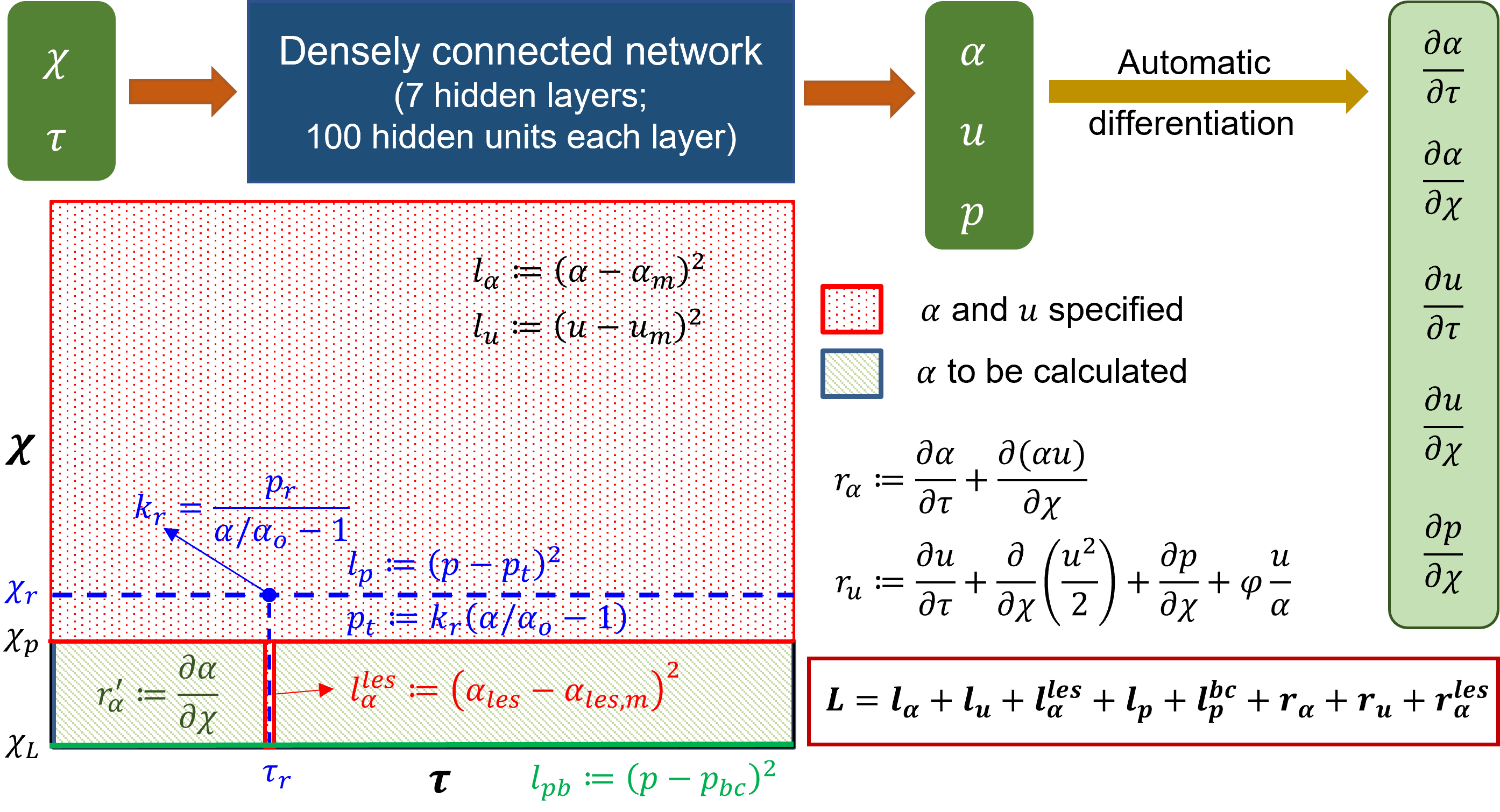}}
  \caption{Details of the physics-informed neural network (PINN). The input and output of the PINN along with the details of the hidden layers are shown at the top. Automatic differentiation was used to calculate the derivative terms for the residuals. The schematic of the domain is shown below. The schematic describes where the different losses were specified.}
  \label{fig-pinn}
\end{figure}
The schematic in Figure \ref{fig-pinn} shows the architecture of the PINN. It takes $\chi$ and $\tau$ as input and predicts $\alpha$, $u$, and $p$. Since the inputs are $\chi$ and $\tau$, automatic differentiation can be used effectively to calculate $\frac{\partial\alpha}{\partial\tau}$, $\frac{\partial\alpha}{\partial\chi}$, $\frac{\partial u}{\partial\tau}$, $\frac{\partial u}{\partial\chi}$, and $\frac{\partial p}{\partial\chi}$ which were used for calculating the terms in Equations \ref{eqn-continuity_nd} and \ref{eqn-momentum_nd}. Aside from the input and the output layers, the PINN consisted of 7 hidden layers with 100 hidden units in each layer. We used $\tanh$ activation function for every layer.

\subsubsection{Losses}
The losses for the PINN consisted of a combination of measurement losses and residuals of the mass and momentum conservation equations. Minimizing the measurement losses ensures that the solutions are consistent with the measurements, and minimizing the residuals ensures that the governing physics behind the problem is followed. Figure \ref{fig-pinn} shows the  locations and time instant at which the different measurement losses and residuals were calculated. As already mentioned in the work-flow, $\alpha$ and $u$ were known at all points proximal to the bolus (marked in red) for all time instants. The measurement losses for $\alpha$ and $u$ for $\chi<\chi_p$ and $0\leq\tau\leq\tau_T$ were as follows:
\begin{align}
l_{\alpha} &=\frac{1}{N_1}\sum_{i=1}^{N_1}\left(\alpha^i - \alpha_m^i\right)^2, \label{eqn-alpha_m} \\
l_u &= \frac{1}{N_1}\sum_{i=1}^{N_1}\left(u^i - u_m^i\right)^2, \label{eqn-u_m} 
\end{align}
wherein the quantities with subscript $m$ represent measured quantities. $\chi_p$ is the proximal end of the LES and $\tau_T$ is the total time (non-dimensional) of bolus transport. Each point $i$ was taken from a Cartesian grid of 99 nodes along $\tau$ and 100 nodes along $\chi$, which leads to $N_1=9900$. Note that we are calling $u_m$ as a measured quantity for the PINN although we calculate it along with $\alpha$ through the interpolated volume $V$ as described in the Section 4.1. This is because the PINN minimizes the square of the difference between prediction of $\alpha$ and $u$ from the network with their already known values (which are analogous to measurements for being already known quantities for the PINN). Additionally, the LES cross-sectional area was known at $\tau=\tau_r$ for $\chi_p<\chi\leq \chi_L$ and the corresponding measurement loss was as follows:
\begin{align}
l_{\alpha}^{les} &=\frac{1}{N_2}\sum_{i=1}^{N_2}\left(\alpha_{les}^i - \alpha_{les,m}^i\right)^2, \label{eqn-alpha_les_m}
\end{align}
wherein $\chi_L$ is the non-dimensional coordinate of the distal end. The points $i$ were taken from a uniform mesh of $N_2=28$ points along $\chi$ at $\tau_r$. 
The measurement loss for pressure was calculated at $\chi=\chi_r$ for $\tau\geq 0$ and was defined as follows:
\begin{align}
l_p = \frac{1}{N_3}\sum_{i=1}^{N_3}\left(p^i - p_t^i\right)^2, \label{eqn-p_m} 
\end{align}
wherein the points $i$ were selected from a uniform mesh of $N_3=98$ along $\tau$ at $\chi=\chi_r$. Additionally, the Dirichlet pressure boundary condition was enforced at $\chi=\chi_L$ for $\tau\geq 0$ through the following loss:
\begin{align}
l_p^{bc}=\frac{1}{N_4}\sum_{i=1}^{N_4}\left(p^i - p_{bc}^i\right)^2, \label{eqn-p_bc} 
\end{align}
wherein $p_{bc}$ is the pressure specified at the distal end of the esophagus and $N_4=99$ with $i$ selected from a uniform grid along $\tau$. 
The residual losses were calculated in the entire domain for $0\leq\chi\leq \chi_L$ and $\tau\geq 0$ according to Equations \ref{eqn-continuity_nd} and \ref{eqn-momentum_nd} as shown below:
\begin{align}
 r_{\alpha}&=\frac{1}{N_5}\sum_{i=1}^{N_5}\left[\frac{\partial \alpha^i}{\partial \tau} + \frac{\partial \left(\alpha^i u^i\right)}{\partial \chi}\right], \label{eqn-alpha_res} \\
 r_u&=\frac{1}{N_5}\sum_{i=1}^{N_5}\left[\frac{\partial u^i}{\partial \tau} + \frac{\partial}{\partial \chi}\left(\frac{(u^i)^2}{2}\right)+\frac{\partial p^i}{\partial \chi} + \varphi\frac{u^i}{\alpha^i}\right], \label{eqn-u_res}
\end{align}
wherein $i$ was randomly sampled from a uniform distribution of points in the entire domain with $N_5=50688$. Finally, the constraint as described in Equation \ref{eqn-les_constraint} led to the following residual:
\begin{align}
r_{\alpha}^{les}=\frac{1}{N_6}\sum_{i=1}^{N_6}\frac{\partial \alpha_{les}^i}{\partial \chi} = 0, \label{eqn-les_res}
\end{align} 
wherein $i$ was randomly sampled from a uniform distribution of points in the domain $[\chi_p,\chi_L]$ and $[0,\tau_T]$ with $N_6=5544$. 
The total loss for the PINN was the sum of all the measurement losses and residuals as follows:
\begin{align}
L=l_{\alpha} + l_u + l_{\alpha}^{les} + l_p + l_p^{bc} + r_{\alpha} + r_u + r_{\alpha}^{les}. \label{eqn_loss}
\end{align}
\begin{figure}[h]
  \centerline{\includegraphics[width=\textwidth]{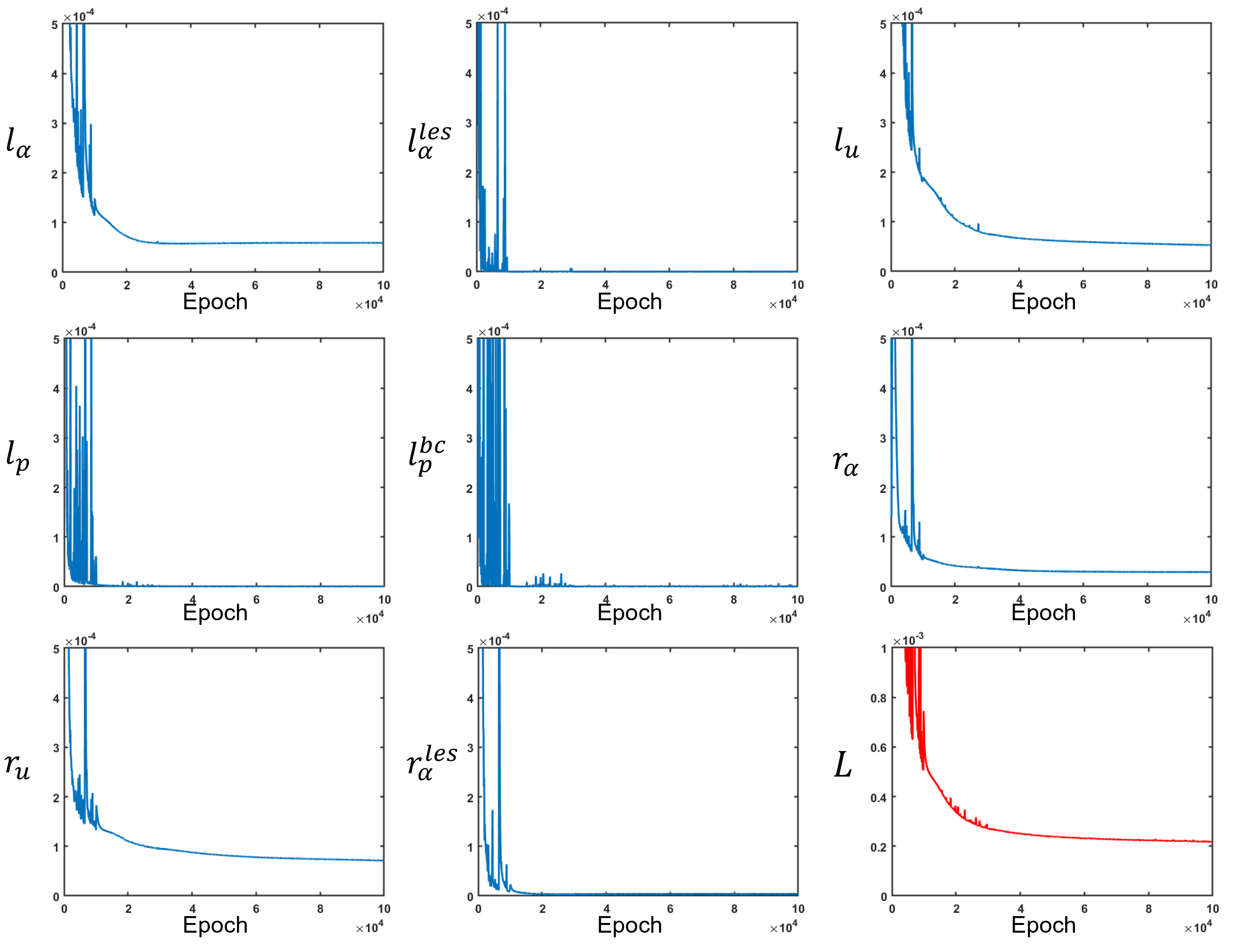}}
  \caption{Measurement losses and residuals along with the total loss. All loss functions were minimized at different rates. The total loss is depicted in red while the other losses are in blue.}
  \label{fig-lr_curves}
\end{figure}

To train the network, the inputs $\chi$ and $\tau$ were normalized with their mean and standard deviation as follows:
\begin{align}
\chi' = \frac{\chi - \mu_{\chi}}{\sigma_{\chi}}, \\
\tau' = \frac{\tau - \mu_{\tau}}{\sigma_{\tau}},
\end{align}
wherein $\mu$ and $\sigma$ are the corresponding mean and standard deviations, respectively for $\chi$ and $\tau$. Hence, the derivatives with respect to $\chi$ and $\tau$ gets modified as follows:
\begin{align}
\frac{\partial}{\partial \chi}\left(\cdot\right) = \frac{1}{\sigma_{\chi}}\frac{\partial}{\partial \chi'}\left(\cdot\right), \\
\frac{\partial}{\partial \tau}\left(\cdot\right) = \frac{1}{\sigma_{\tau}}\frac{\partial}{\partial \tau'}\left(\cdot\right).
\end{align} 

\subsubsection{Training}
The network was trained using Tensorflow \cite{tensorflow} for 100000 epochs. We used an Adam \cite{adam} optimizer to minimize the losses. A piecewise constant decayed learning rate was used to minimize the losses efficiently. The learning rate was 0.001 for the first 10000 epochs, 0.0001 for the next 20000 epochs, and 0.00003 for the last 70000 epochs. The final values for $l_{\alpha}$, $l_u$, $l_{\alpha}^{les}$, $l_p$, $l_p^{bc}$, $r_{\alpha}$, $r_u$, $r_{\alpha}^{les}$ were $5.9\times10^{-5}$, $9.8\times10^{-7}$, $5.3\times10^{-5}$, $4.0\times10^{-7}$, $2.7\times10^{-7}$, $2.9\times10^{-5}$, $7.2\times10^{-5}$, and $3.8\times10^{-6}$, respectively. Figure \ref{fig-lr_curves} shows the learning curves for the various loss functions. The final total loss was $2.2\times10^{-4}$.

\subsection{Esophageal wall stiffness and active relaxation}
The esophageal wall stiffness and active relaxation were calculated as described in Halder et al. \cite{halder2021}. A few manipulations of Equation \ref{eqn-tube_law} yields the following:
\begin{align}
\frac{P-P_o}{\frac{A}{A_o}-1} = \frac{K}{\theta}\left[1 - \frac{\theta-1}{\frac{A}{A_o}-1}\right]. \label{eqn-tube_law_manipulation}
\end{align}
The active relaxation parameter $\theta$ is always greater than 1 at the location of the bolus. Additionally, $A>A_o$ at the bolus due to the distension of esophageal walls. Thus, the second term of Equation \ref{eqn-tube_law_manipulation} is always greater than 0. Using these constraints we arrive at the following inequality:
\begin{align}
\frac{K}{\theta} \geq \frac{P-P_o}{\frac{A}{A_o}-1}, \label{eqn-stiffness}
\end{align}
where $K/\theta$ estimates the lower bound of the esophageal stiffness and incorporates the effect of active relaxation as well. The active relaxation of the esophageal walls was estimated as follows:
\begin{align}
\theta = \frac{\alpha}{\alpha_r}, \label{eqn-theta}
\end{align}
where $\alpha_r$ is the reference non-dimensional cross-sectional area near the distal end of the esophageal body at $\chi=\chi_r$ as shown in Figure \ref{fig-problem_definition}. The value of the $\theta$ at $\chi_r$ was assumed to be 1 and acted as a reference to calculate active relaxation for all $\chi<\chi_r$.
 
\section{Results and discussion}
\begin{figure}[h]
\captionsetup[subfigure]{justification=centering}
\begin{subfigure}[c]{.49\textwidth}
  \centering
  \includegraphics[scale=0.55]{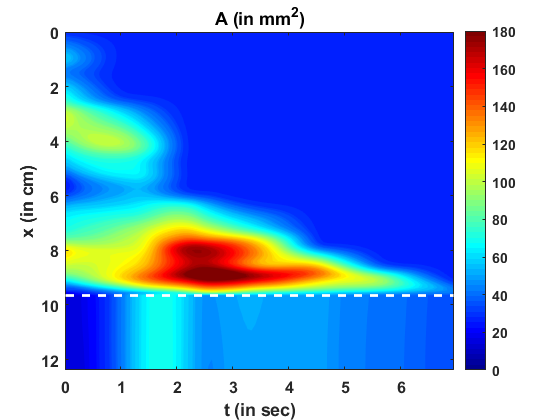}
  \caption{}
  \label{fig-A}
\end{subfigure}
\begin{subfigure}[c]{.49\textwidth}
  \centering
  \includegraphics[scale=0.55]{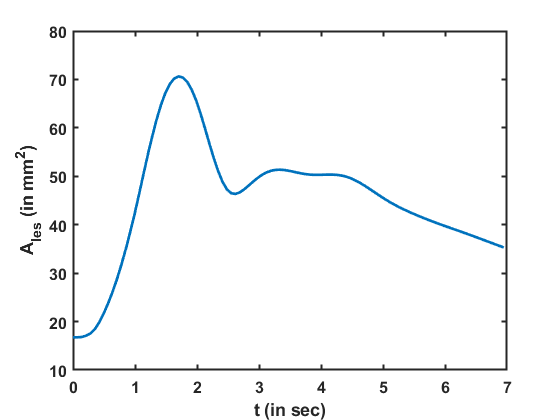}
  \caption{}
  \label{fig-A_les}
\end{subfigure}
\caption{Variation of cross-sectional area as predicted by the PINN. (a) Variation of $A$ as a function of $x$ and $t$. The dashed white line indicates the proximal end of the LES. The cross-sectional area above the dashed line was known from the MRI and its prediction by the PINN was ensured by minimizing Equation \ref{eqn-alpha_m}. There is no variation of $A$ along $x$ within the LES due to the constraint described by Equation \ref{eqn-les_constraint}, (b) Variation of the LES cross-sectional area as a function of time. It had the greatest magnitude near the instant that the LES was visible in the MRI.} 
\end{figure}

The PINN predicts the non-dimensional cross-sectional area, fluid velocity, and fluid pressure by minimizing a set of measurement losses as well as ensuring that the physics of the fluid flow problem is followed throughout. The variation of the predicted cross-sectional area (in its dimensional form) is shown in Figure \ref{fig-A} for all values of $x$ and $t$. The values of the cross-sectional areas inside the bolus proximal to the LES were obtained from measurements and their prediction was based on the minimization the measurement loss as described by Equation \ref{eqn-alpha_m}. 

\begin{figure}[h]
\captionsetup[subfigure]{justification=centering}
\begin{subfigure}[c]{.49\textwidth}
  \centering
  \includegraphics[scale=0.45]{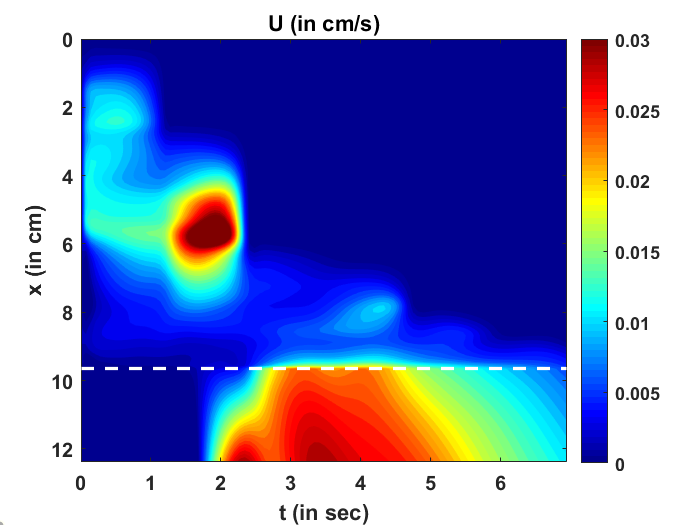}
  \caption{}
  \label{fig-U}
\end{subfigure}
\begin{subfigure}[c]{.49\textwidth}
  \centering
  \includegraphics[scale=0.45]{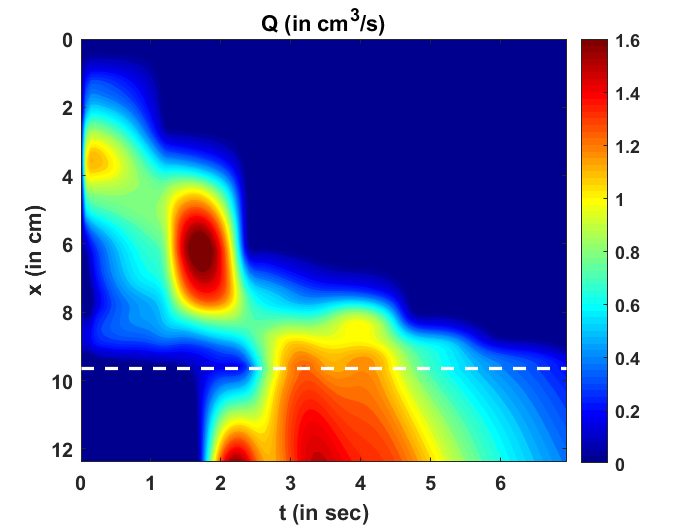}
  \caption{}
  \label{fig-Q}
\end{subfigure}
\caption{Variation of velocity and flow rate. (a) Variation of $U$ as predicted by the PINN. There are two high velocity zones: one at $x~6$ cm, $t=2$ s and the other at the LES for $t>2$ s. These high velocity zones match the regions of low cross-sectional areas: (b) Variation of the mean flow rate calculated as $Q=AU$. The high flow rate matches the high velocity zones, but there is a smoother transition of $Q$ at the proximal end of the LES compared to $U$.} 
\end{figure}

The cross-sectional areas proximal to the bolus cannot be visualized in MRI because the fluid contrast media was completely displaced by the peristaltic contraction and the dynamic MR imaging cannot distinguish the esophagus from surrounding tissue. Hence, we assigned the inactive cross-sectional area $A_o$ to the esophagus proximal to the bolus. We found that this assignment does not impact the prediction of any of the physical quantities using PINN. This is because the velocity (and flow rate) proximal to the bolus is automatically predicted as zero (as shown in Figures \ref{fig-U} and \ref{fig-Q}) with this assignment, and since the pressure boundary condition is specified at the distal end, the pressure calculation inside the domain does not depend on the behavior proximal to the bolus. The variation of LES cross-sectional area can be seen below the dashed line in Figure \ref{fig-A}. The LES cross-sectional area does not vary along $x$ and only varies along $t$. This is because we enforced the constraint as described in Equation \ref{eqn-les_constraint}.

The variation of the LES cross-sectional area is shown more clearly in Figure \ref{fig-A_les}. The prediction of $A_{les}$ depends on the reference LES cross-sectional area observed at a single time instant, the conservation laws, and the reference pressure prediction at $\chi_r$. $A_{les}$ has the greatest magnitude near the instant when the LES cross-sectional area was observed in the MRI and has lesser values farther away from that instant. This matches our observation from the MRI images that the LES could not be visualized most of the time. Hence, since the effectiveness of esophageal transport essentially depends on how effectively the esophagus empties, the LES cross-sectional area is an important physiomarker of esophageal function. Greater LES cross-sectional area facilitates esophageal emptying while it is becomes unnecessary for the LES to have large cross-sectional area when the bolus has almost completely emptied. Similar LES behavior is evident in Figure \ref{fig-A_les} where it was greater during the emptying process and minimal when bolus emptying was nearly complete.

The variation of bolus fluid velocity and flow rate are shown in Figures \ref{fig-U} and \ref{fig-Q}, respectively. It can be seen that there are two major high-velocity zones. The first high-velocity zone is near $x=6$ cm and $t=2$ sec. Comparing this region with the Figure \ref{fig-A} it is evident that the cross-sectional area at that location and time was less than at its adjacent regions. The second high-velocity zone was in the LES. This corresponds to low cross-sectional area as well. Thus, the velocities are greater at lower cross-sectional areas which is intuitive for low viscosity fluids. The flow rate is the rate at which the bolus is emptied out of the esophagus and zones with high flow rate are similar to those with high-velocity. However, there is a smoother transition of flow rate from the esophageal body to the LES compared to the velocity field. This is because the LES cross-sectional area was much smaller than that of the esophageal body requiring that the fluid velocity needed to increase more to maintain the same flow rate.

\begin{figure}[H]
  \centerline{\includegraphics[scale=0.6]{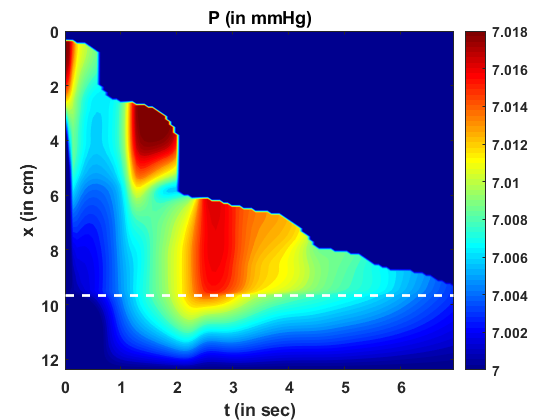}}
  \caption{Variation of pressure as a function of $x$ and $t$. Two major high-pressure zones can be identified wherein the fluid locally accelerates, making the corresponding fluid velocity greater in those regions. Note that the magnitude of dynamic pressure variations are minimal compared to the total pressure.}
  \label{fig-P}
\end{figure}

The variation of fluid pressure is shown in Figure \ref{fig-P}. The pressure gradients along $x$ drives the fluid through the esophagus. On comparing Figures \ref{fig-U} and \ref{fig-P}, we can see that the high-pressure gradients match the high-velocity zones. This is because the high-pressure gradients locally accelerate the fluid. Note that the pressure variations are minimal compared to the magnitudes of the pressure. An intragastric pressure of 7 mmHg was used as a boundary condition for pressure at the distal end which is in the normal range for a healthy subject. The thoracic pressure was assumed to be 0 mmHg. Thus, the intrabolus pressure must be greater than the intragastric pressure to empty into the stomach. The major portion of this pressure ($\sim$7 mmHg) is developed by the elastic distention of the esophageal walls. A small portion of the total intrabolus pressure ($\sim$0.01 mmHg) is attributable to the local acceleration or deceleration of the bolus fluid. Since the MRI shows only the movement within an already distended esophagus, the calculated pressure variations are minimal and correspond to local acceleration or deceleration of the fluid. This observation regarding dynamic pressure variations was also observed in mechanics-based analysis of fluoroscopy \cite{halder2021}. 

The total intrabolus pressure as shown in Figure \ref{fig-P} is within the normal range according to CCv4.0, leading us to conclude that our specifications of the intragastric pressure and thoracic pressure were valid. The prediction of $A_{les}$ depends on the pressure gradients and not the actual magnitude of pressure. Therefore, the prediction of $A_{les}$ remains the same irrespective of the boundary condition chosen for $P$. Figures \ref{fig-A_les}, \ref{fig-U}, \ref{fig-Q}, and \ref{fig-P} also point at an important feature of the LES. The greatest LES cross-sectional area (at approximately 1.8 s) neither match the greatest pressure nor the greatest velocity (or flow rate) across the LES. This demonstrates that the LES opening is not governed passively by intrabolus pressure. If the LES was passively opened by elastic distention due to the intrabolus pressure, then the maximum LES cross-sectional area would coincide with the maximum pressure gradient. Since that is not observed, it can be concluded that the LES cross-sectional area also involves neuromuscular relaxation.
\begin{figure}[H]
  \centerline{\includegraphics[scale=0.6]{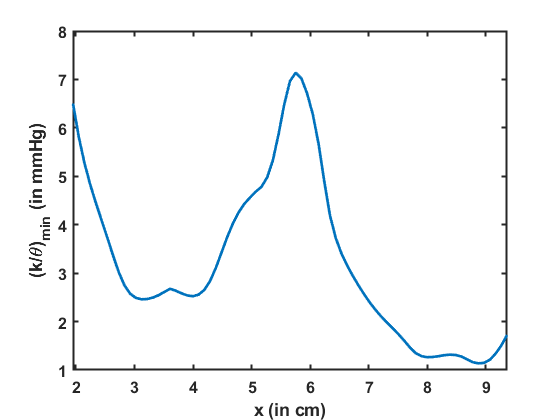}}
  \caption{Variation of the minimum esophageal wall stiffness along the length of the esophagus. This measure of stiffness accounts for active relaxation and captures the wall characteristics when the esophagus was distended. The stiffness is shown only for the esophageal body proximal to the LES.}
  \label{fig-kmin}
\end{figure}

Esophageal wall stiffness (along with the effect of active relaxation) was estimated by the parameter $K/\theta$. The minimum value of $K/\theta$ corresponds to the lower bound of the effective stiffness of the esophageal walls when distended. Since the cross-sectional area of the esophagus is not visible in MRI, any prediction regarding the stiffness at those locations would be inaccurate. Hence, predictions of wall stiffness can only be made at regions where the esophagus is distended i.e., at the location of the bolus. However, the distended esophageal walls also undergo active relaxation to accommodate an incoming bolus as well as minimize intrabolus pressure. The combined behavior of the passive elastic distention of the esophageal walls and active relaxation is captured by the parameter $K/\theta$. Since $K/\theta$ as described by Equation \ref{eqn-stiffness} estimates the lower bound of the effective esophageal stiffness, the most accurate estimate of $K/\theta$ occurs when the esophageal walls are most distended. The maximum distension corresponds to the minimum value of $K/\theta$, which is shown in Figure \ref{fig-kmin}. The minimum $K/\theta$ at each $x$ was calculated for all values of $t$. Note that the high value of $(K/\theta)_{min}$ near $x=6$ cm in Figure \ref{fig-kmin} matches with the low cross-sectional area region in Figure \ref{fig-A}. This makes sense because the esophagus would distend less at locations of greater stiffness. It should be noted that although the stiffness appears high at $x=6$ cm, it is does not necessarily mean that the esophageal tissue is stiffer at that location. When the esophageal wall comes in contact with surrounding organs, it appears stiffer due to the effect of those organs on the esophagus. Since, all calculations are made using only bolus geometry, it is impossible to distinguish the effect of other organs outside the esophagus. Hence, we hypothesize that the lower values of $(K/\theta)_{min}$ estimate the true stiffness of the esophageal walls and the greater value of $(K/\theta)_{min}$ near $x=2$ cm is likely a composite measure partly attributable to extrinsic compression. Close to the advancing peristaltic contraction, $\theta<1$ so, $(K/\theta)_{min}$ takes a greater value and the esophagus seems to be locally stiffer. Also, note that we have not included the EGJ in Figure \ref{fig-kmin}. This is because we did not define the problem with the tube law applied at the EGJ because applying the tube law at the EGJ would not result in a unique solution. The mechanical properties of the esophageal walls have been estimated experimentally in Orvar et al. \cite{orvar1993}, Patel and Rao \cite{patel1998}, and  Kwiatek et al. \cite{kwiatek2010}. In those studies, the esophagus was distended and the cross-sectional area and the pressure developed inside were recorded. A straight line was fitted to quantify the linear relationship between cross-sectional area and pressure in an inactive esophagus. The slope of the line measured the quantity $A_o/K$ which was in the range 9.1-11.6 $\mathrm{mm}^2/\mathrm{mmHg}$. Using the typical range of $A_o$ as described in Xia et al. \cite{xia2009}, i.e. $7-59$ $\mathrm{mm}^2$, the stiffness of the esophageal walls was found to lie in the range $0.6-6.5$ mmHg. The  effective stiffness as shown in Figure \ref{fig-kmin} lay in the range $1-7$ mmHg, which is of the same order of magnitude as observed in the other studies.

\begin{figure}[ht]
  \centerline{\includegraphics[scale=0.6]{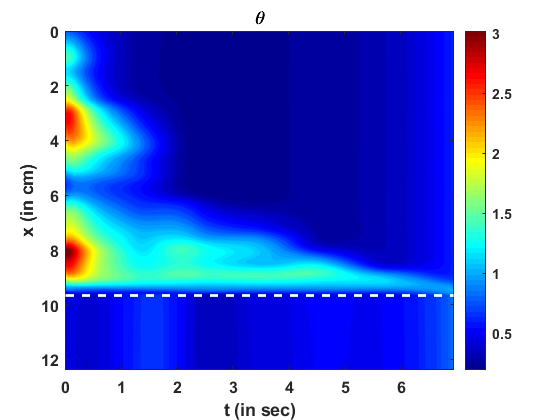}}
  \caption{Variation of active relaxation as a function of $x$ and $t$. The dashed line corresponds to the proximal end of the LES. Since the tube law was not specified at the LES, the active relaxation is meaningful only in the esophageal body (above the dashed line).}
  \label{fig-theta}
\end{figure}

The parameter $\theta$ quantifies the amount of active relaxation of the esophageal walls to facilitate distention, and consequently, decrease the local intrabolus pressure and increase the flow rate. The variation of the active relaxation parameter $\theta$ is shown in Figure \ref{fig-theta}. As described by Equation \ref{eqn-theta} and comparing Figure \ref{fig-theta} and \ref{fig-A}, it is evident that the locations of the high values of $\theta$ match the location of the high values of $A$, and, similarly lower values of $\theta$ match the lower values of $A$. Note that $\theta$ quantifies the active relaxation in the esophageal body and not the LES. Comparing Figures \ref{fig-kmin} and \ref{fig-theta} shows that locations of greater stiffness correspond to locations of lower active relaxation and vice-versa. Similar to $(K/\theta)_{min}$ as described above, the impact of tissues and organs outside the esophagus impacts the prediction of $\theta$ as well. Hence, the low value of $\theta$ near $x=6$ cm does not necessarily mean a lack of active relaxation, but most likely the influence of structures outside the esophagus. Hence, we hypothesize that the greater values of active relaxation are closer to the actual active relaxation of the esophageal walls.

\subsection*{Limitations}
Although MRI-MECH provides valuable insights about the nature of transport and the mechanical state of the esophagus, it has limitations. Currently, manual segmentation of the bolus geometry is more accurate and is reasonable for a low temporal resolution of the dynamic MRI, but can become tedious with improved temporal resolution. Automatic segmentation using deep learning techniques might be helpful in that aspect, but also increases the risk of inaccurate segmentation without a large training dataset. Bolus transport as visualized in MRI provides no information proximal to the bolus (a similar problem occurs in fluoroscopy as well). Hence, the MRI-MECH cannot predict anything meaningful proximal to the bolus. Thus, MRI-MECH cannot be used to estimate the contraction strength, for which other diagnostic techniques should be used such as HRM or FLIP. The esophageal wall properties and neurally-activated relaxation were estimated solely through the bolus shape and movement. But the bolus shape and movement depend not only on the esophageal walls but also on the impact of organs surrounding the esophagus. This is a limitation of the MRI-MECH framework in the predicting the state and functioning of the esophagus due to lack of information about the impact of the surrounding organs. Finally, the prediction of intrabolus pressure and the esophageal wall stiffness depends on the specification of the correct intragastric pressure. This becomes a limitation for MRI-MECH since the intragastric pressure is not known in the MRI, and so, we used a reference value from literature. Accurate measurement of the intragastric pressure through other diagnostic techniques such as HRM will increase the accuracy of the MRI-MECH predictions of intrabolus pressure and wall stiffness.

\section{Conclusion}
We presented a framework called MRI-MECH that uses dynamic MRI of a swallowed fluid to quantitatively estimate the mechanical health of the esophagus. The bolus geometry, which tracks the inner cross-section of the esophagus, was extracted through manual segmentation of the MR image sequence and was used as input to the MRI-MECH framework. MRI-MECH modeled the esophagus as a one-dimensional flexible tube and used a physics-informed neural network (PINN) to predict the fluid velocity, intrabolus pressure, esophageal wall stiffness, and active relaxation. The PINN minimized a set of measurement losses to ensure that the predicted quantities matched the measured quantities, and a set of residuals to ensure that the physics of the fluid flow problem was followed, specifically, the mass and momentum conservation equation in one-dimension. The LES cross-sectional area is very difficult to visualize in MRI because it is significantly smaller than the cross-sectional area at the esophageal body. In this regard, MRI-MECH enhances the capability of the dynamic MRI by calculating the LES cross-sectional area during the esophageal emptying. We found that our predictions of the intrabolus pressure and the esophageal wall stiffness match those reported in other experimental studies. Additionally, we showed that the dynamic pressure variations that occur because of local acceleration/deceleration of the fluid were negligible compared to the total intrabolus pressure, whose main contribution was the elastic deformation of the esophageal walls. The mechanics-based analysis with detailed three-dimensional visualization of the bolus in MRI leads to significantly better prediction of the state of the esophagus compared to two-dimensional X-ray imaging such as esophagram and fluoroscopy, and can be easily extended to other medical imaging techniques such as computerized tomography (CT). Thus, MRI-MECH provides a new direction in mechanics-based non-invasive diagnostics that can potentially lead to improved clinical diagnosis.

\section*{Acknowledgments}
This research was supported in part through the computational resources and staff contributions provided for the Quest high performance computing facility at Northwestern University which is jointly supported by the Office of the Provost, the Office for Research, and Northwestern University Information Technology.

\section*{Funding Data}
\begin{itemize}
\item National Institutes of Health (NIDDK grants DK079902 \& DK117824; Funder ID: 10.13039/100000062)
\item National Institutes of Health (NCATS grant TL1TR001423)
\item National Science Foundation (OAC grants 1450374 \& 1931372; Funder ID: 10.13039/100000105)
\end{itemize}

\bibliographystyle{elsarticle-num}
\bibliography{mri_mech}
\newpage

\section*{Appendix}
\setcounter{figure}{0}
\makeatletter 
\renewcommand{\thefigure}{\@Roman\c@figure}
\makeatother
\subsection*{MRI segmentation using deep learning}
In this work, we also demonstrate the use of a convolutional neural network (CNN) called a 3D U-Net to perform semantic segmentation based on a limited number of sample 2D slices of the 3D MRI image. The 3D U-Net is built on the U-Net, which is composed of a contracting encoder path and an expansive decoder path. A key differentiating feature of the 3D U-Net is that instead of operating on the 2D images that the U-Net expects as input, the former operates on volumetric (3D) data. Automatic segmentation using the 3D U-Net was viable due to the repetitiveness of the structures and variations of the input images, especially for volumetric biomedical data. 

The network architecture used was the same as the 3D U-Net, with the only change being that the number of channels is 1, instead of 3. Batch normalization was performed after each convolutional layer to prevent overfitting. The network had a total of 16,324,929 parameters, of which 4,672 were non-trainable and 16,320,257 were trained.

The 3D U-Net was trained using 5 training sets, with each consisting of an image and a mask. The original MRI scans are composed of folders for each time instant, which contain a series of 64 .dcm files representing each slice. MRIcron, a NIfTI image viewer that contains the DICOM to NIfTI converter dcm2nii, was used to convert these raw images into .nii files to be loaded into the model. To prepare the training masks, we used ITK-SNAP to perform segmentation via manual delineation, where voxels in the images were given a label of 1 if they were determined to be part of a bolus, and 0 if they were not. For the benchmark case, the images were manually segmented with intervals of 2, 4, and 2 slices respectively in the $x$, $y$, and $z$ directions. In order to determine the effects of the sparsity of the annotation on the training of the model as well as the accuracy of the prediction, manual segmentations were made on the same set of images using the same method, except with intervals in the $x$, $y$, and $z$ directions of (4, 8, 4), (4, 8, 8), in addition to a training set with only one segmentation slice in each direction to represent a more extreme case of sparse annotation. Another image/mask set was used for validation during training, and the 7th set was for testing to compare prediction accuracy among the different cases. The amount of data used in the training of the model is adequate, as in many biomedical volumetric image classification situations, only two images are required to attain reasonable accuracy, along with a weighted loss function and data augmentation. 

The original images and masks were of size $160 \times 160 \times 64$, but we implemented a cropping procedure along the $x$-direction such that the eventual inputs to the neural network were of size $160 \times 80 \times 64$. This was done in an endeavor to reduce the number of voxels in the input image and thereby reduce processing time, whilst still retaining the area of focus – the esophagus and bolus. Data augmentation was also implemented using techniques such as Gaussian blur, image sharpening, random variation of image brightness, contrast normalization, and elastic deformation.

For our segmentation problem, we used the weighted Sorensen Dice Coefficient to measure the similarity between the predictions and the training set. The weighted dice coefficient loss was used as the loss function during training.

\begin{figure}[h]
    \centering
    \includegraphics[width=0.8\textwidth]{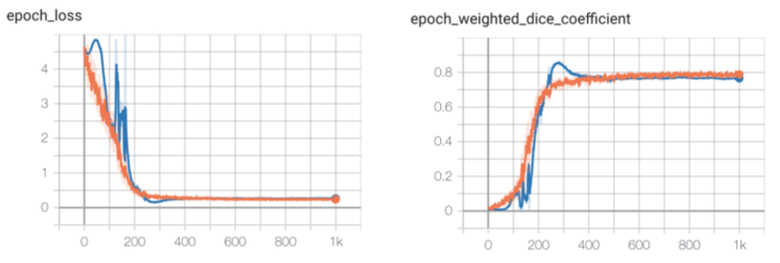}
    \caption{Loss and weighted dice coefficient during training for the benchmark case. The curve for the training set is in orange, and blue represents the validation set.}
    \label{fig-app_learning_curves}
\end{figure}

\begin{figure}[h!]
    \centering
    \includegraphics[width=0.5\textwidth]{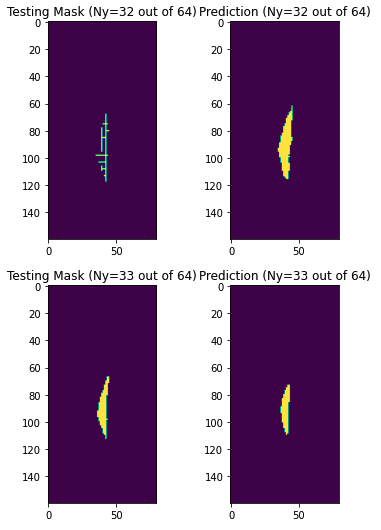}
    \caption{Side-by-side comparison of the testing mask (annotated manually) vs. the prediction for the benchmark case for some sample $y$ locations.}
    \label{fig-app_prediction}
\end{figure}

Several hyperparameter tuning trials were conducted to find the optimal hyperparameters and conditions for training the model. Through experimentation, we found that training for 1000 epochs with an initial learning rate of $10^{-2}$ using a learning rate scheduler to decrease the learning rate by a factor of 3 every 250 epochs proved to be the most reliable. The training was conducted using Keras, a high level neural network API of TensorFlow, and the optimization algorithm used was stochastic gradient descent (SGD). The learning curves are shown in Figure \ref{fig-app_learning_curves}. A maximum weighted dice coefficient of 0.8575 was attained at epoch 282 during training. The predictions were converted to their binary forms with voxels with values greater than 0.05 classified as 1 (bolus), and those equal to or below 0.05 is set to 0 (background). Figure \ref{fig-app_prediction} shows the predicted segmentation.

Retaining the same hyperparameters, we also ran trials with different levels of annotation sparsity. Let us denote the number of slices in the $x$, $y$, and $z$ directions as $(N_x,N_y,N_z)$. The benchmark trial had (2,4,2), and saw a final validation weighted dice coefficient of 0.7679. The final weighted dice coefficient for the (4,8,8), (4,8,4), and (1,1,1) cases were respectively 0.4608, 0.5289, and 0.3655. Thus, in this trial, we found that the benchmark case of (2,4,2) yielded the highest accuracy.

Ultimately, the training of a 3D U-Net and its application for the ascertainment of the location and shape of the bolus was to evaluate the potential of using the output data for MRI-MECH. In this regard, there may be room for further work, particularly in the post-processing of the prediction since more accuracy in the prediction is desirable for mechanics-based simulations.
\end{document}